\documentclass[preprint,aps,nofootinbib]{revtex4}

\usepackage{graphicx}

\usepackage{amsmath}
\usepackage{amsfonts}
\usepackage{amssymb}
\usepackage{color}

\def\be{\begin{equation}}
\def\ee{\end{equation}}
\def\bea{\begin{eqnarray}}
\def\eea{\end{eqnarray}}

\def\slashchar#1{\setbox0=\hbox{$#1$}           
   \dimen0=\wd0                                 
   \setbox1=\hbox{/} \dimen1=\wd1               
   \ifdim\dimen0>\dimen1                        
      \rlap{\hbox to \dimen0{\hfil/\hfil}}      
      #1                                        
   \else                                        
      \rlap{\hbox to \dimen1{\hfil$#1$\hfil}}   
      /                                         
   \fi}

\begin{document}

\vspace*{1cm}

\title{Adler's Zero and Effective Lagrangians for Nonlinearly Realized Symmetry}

\author{\vspace{0.5cm} Ian Low}
\affiliation{\vspace{0.5cm}
\mbox{High Energy Physics Division, Argonne National Laboratory, Argonne, IL 60439}\\
\mbox{Department of Physics and Astronomy, Northwestern University, Evanston, IL 60208} \\
 \vspace{0.5cm}
}

\begin{abstract}
\vspace{1cm}
Long ago Coleman, Callan, Wess and Zumino (CCWZ) constructed the general effective lagrangian for nonlinearly realized symmetry by finding all possible nonlinear representations of the broken group $G$ which become linear when restricted to the unbroken group $H$. However, in the case of  a single Nambu-Goldstone boson (NGB), which corresponds to a broken $U(1)$, the effective lagrangian can also be  obtained by  imposing a constant shift symmetry. In this work we generalize the shift symmetry approach to multiple NGBs and show that, when they furnish a linear representation of $H$ that can be embedded in a symmetric coset, it is possible to derive the CCWZ lagrangian by imposing 1) the "Adler's zero condition," which requires scattering amplitudes to vanish when emitting a single soft NGB, and 2) closure of shift symmetry with the linearly realized symmetry; knowledge of the broken group $G$ is not required at all. Using only  generators of $H$,  the NGB covariant derivative and the associated gauge field can be computed to all orders in the NGB decay constant $f$.

\end{abstract}

\maketitle

\section{Introduction}
\label{sect:intro}

Effective lagrangians for nonlinearly realized symmetry were first introduced by Gell-Mann and L\'{e}vy in Ref.~\cite{GellMann:1960np} in the context of $SU(2)_L\times SU(2)_R$ chiral symmetry breaking in 1960, and later was found to play an essential role in many areas of theoretical physics. In a pair of elegant papers, CCWZ presented a general method to construct such effective lagrangians, by starting from a broken group $G$ in the UV and considering all possible nonlinear representations of $G$ which become linear when restricted to the unbroken subgroup $H$ \cite{Coleman:sm,Callan:sn}. The NGBs then parameterize the coset space $G/H$, giving rise to the familiar counting rule that the number of NGBs equals the number of broken generators.

When there is only one NGB, corresponding to a broken $U(1)$ group, CCWZ gives a particular simple form of effective lagrangian,
\be
\frac12 \partial_\mu \pi \partial^\mu \pi + {\cal O}(\partial^4\pi^2) \ ,
\ee 
where the NGB has no potential and is always derivatively coupled. Alternatively, the above lagrangian can also be derived by imposing a constant shift in the NGB field, $\pi\to \pi + \epsilon$, which forbids any non-derivatively coupled terms in the lagrangian. The derivative coupling, or the shift symmetry, manifests itself in the $S$-matrix elements as the Adler's zero \cite{Adler:1964um}, which states the scattering amplitudes of NGBs must vanish when emitting a single soft NGB.

In this work we will generalize the shift symmetry approach to multiple NGBs furnishing a linear representation of a simple Lie group $H$, by imposing only the Adler's zero condition and closure of the shift symmetry with the linearly realized $H$ symmetry. This does not always happen, and we derive the Closure Condition when the effective lagrangian can be obtained in this way. The Closure Condition turns out to be equivalent to requirement that the representation can be embedded in a symmetric coset $G/H$. In the end, we find the full CCWZ lagrangian can be reconstructed using only generators of the unbroken group $H$, without ever referring to any broken group $G$ in the UV.

Our finding suggests interactions of NGBs are determined by their transformation property under the group $H$ in the IR, and are independent of the broken group $G$ in the UV. This result has  important implications in phenomenological models where the Higgs boson arises as a pseudo-NGB and will be explored elsewhere \cite{minimalhiggs}.

This work is organized as follows. In the next section we provide a lightning review of the CCWZ formalism, followed by a discussion on the Adler's zero condition and the shift symmetry for a single flavor of NGB. Then in Section \ref{sect:u1} we generalize the shift symmetry to the case of two NGBs transforming under an unbroken $U(1)$ as a complex scalar, while in Section \ref{sect:general} we consider the general shift symmetry for a simple Lie group, deriving both the Closure Condition and simple formulas for the NGB covariant derivatives and the associated gauge field. Finally we conclude and provide some discussions in Section \ref{sect:conclusion}.

\section{A Brief Overview of CCWZ}
\label{sec:ccwz}

We will use the notation $T^i$ for the unbroken group generators in $H$ and $X^a$ the broken generators in $G/H$. In general they have the following commutation relations
\be
[T^i, T^j] = i  f^{ijk} T^k \ , \quad [T^i, X^a] = i f^{iab} X^b \ , \quad [X^a, X^b]= i f^{abi} T^i + i f^{abc} X^c \ .
\ee
The first set of commutators, $[T^i, T^j]$, states that $H$ is a subgroup of $G$, while the second set of commutators, $[T^i, X^a]$,  imply that the broken generators $X^a$ furnish a linear representation of the algebra of $H$. For symmetric coset $f^{abc}=0$ and $[X, X]\sim T$.

The NGB is parametrized by the matrix
\begin{equation}
\label{eq:nonli}
 \xi= e^{i {\Pi}/f}, \quad \Pi=  \pi^a X^a ,
\end{equation}
where $f$ is the analogue of the pion decay constant in QCD chiral lagrangian. Notice that $\xi\in G/H$ and, under the action of a group element $g\in G$, transforms as 
\be
\label{eq:gxixiu}
g \ \xi = \xi^\prime\, U(g, \xi)\ , \quad U(g,\xi) \in H \ .
\ee
The Goldstone-covariant derivative ${\cal D}_\mu$  and the associated gauge field ${\cal E}_\mu$ are derived from the Cartan-Maurer one-form \cite{Coleman:sm,Callan:sn}:
\begin{eqnarray}
\label{eq:cartan}
\xi^\dagger \partial_\mu \xi  &=& i {\cal D}_\mu ^a X^a + i {\cal E}_\mu^i T^i \equiv i {\cal D}_\mu + i {\cal E}_\mu .
\end{eqnarray}
The Goldstone-covariant derivative transforms homogeneously under the nonlinearly realized symmetry transformation, while the associated gauge field transforms non-homogeneously:
\be
{\cal D}_\mu \to U {\cal D}_\mu U^{-1} \ , \quad {\cal E}_\mu \to U {\cal E}_\mu U^{-1} - (\partial_\mu U) U^{-1} \ .
\ee
Then the  two-derivative interaction in the Lagrangian is given by
\begin{equation}
\label{lag}
{\cal L}_2= \frac{f^2}2 \text{Tr} ({\cal D}^\mu {\cal D}_\mu) \ ,
\end{equation}
while the associated gauge field can be used with ordinary derivative to construct a covariant derivative, $\partial_\mu + {\cal E}_\mu$, for matter fields furnishing a linear representation of  $H$.

As a demonstration, let's apply the CCWZ to the coset $SU(2)/U(1)$, where the are two NGBs: $\phi^1$ and $\phi^2$. We will choose a basis 
\be
X^a = \frac1{\sqrt{2}}\sigma^a \ ,\ \  a=1,2,  \qquad T=\frac1{\sqrt{2}}\sigma^3 \ ,
\ee
where $\{\sigma^1, \sigma^2, \sigma^3\}$ are the Pauli matrices and the group generators are normalized to ${\rm Tr}(X^a X^b)=\delta_{ab}$. The NGB matrix can be written in terms of a complex scalar $\phi=(\phi_1+i\phi_2)/\sqrt{2}$
\be
\xi = e^{i \Pi/f} \ , \quad \Pi=\sum_{a=1,2} \phi^a X^a =\left(
\begin{array}{cc}
 0 & \phi \\
 \phi^* & 0 
 \end{array} \right) \ .
\ee 
The two-derivative lagrangian following from Eq.~(\ref{lag}) is 
\bea
\label{eq:su2u1}
{\cal L}_2 &=& |\partial_\mu \phi|^2-\frac1{3f^2} \left| \phi^*\partial_\mu\phi -\phi\partial_\mu \phi^*\right|^2 + \frac{8}{45f^4}\left| \phi^*\partial_\mu\phi -\phi\partial_\mu \phi^*\right|^2 |\phi|^2\nonumber \\
&& \qquad - \frac{16}{315f^6} \left| \phi^*\partial_\mu\phi -\phi\partial_\mu \phi^*\right|^2 |\phi|^4 + \cdots \ .
\eea
Obviously $SU(2)/U(1)$ is not the only coset containing a charged NGB. Next we consider a  more complicated coset, the well-known $SU(5)/SO(5)$ coset from the littlest Higgs model introduced in Ref.~\cite{ArkaniHamed:2002qy}, which considered a nonlinear sigma model arising from a $5\times 5$ symmetric matrix $\Phi$, transforming under the
global $SU(5)$ symmetry as $\Phi \to V \Phi V^T$, with a vacuum expectation value
\begin{equation}
\Sigma_0 = \left( \begin{array}{ccc}
                     &      & \ \openone\  \\
                     &   1  &          \\
           \ \openone\  &      &     
                  \end{array}     \right)\ .
\end{equation}
$\Sigma_0$ breaks $SU(5) \to SO(5)$.  The NGB fields, $\Pi = \pi^a X^a$, are parametrized as 
\begin{equation}
\Sigma=  e^{i 2\Pi/f} \Sigma_0\  ,
\end{equation}
where $\Pi$ contains a doublet under a $SU(2)$ subgroup of the $SO(5)$:
\begin{equation}
\Pi = \left( \begin{array}{ccc}
            \phantom{\sigma} & \frac{H}{\sqrt{2}} & \phantom{\phi} \\
             \frac{H^\dagger}{\sqrt{2}}  &   & \frac{H^T}{\sqrt{2}}  \\
            \phantom{\phi^\dagger}     &  \frac{H^*}{\sqrt{2}}  &   
             \end{array}  \right),
\end{equation}
In components the doublet scalars are $H = (h^+\ h^0)^T$.  Focusing on the upper component, $h^+$, the two-derivative interactions are \cite{ArkaniHamed:2002qy},\footnote{We have ignored the gauge fields which are irrelevant for our purpose.} 
\bea
\label{eq:su5so5}
{\cal L}_2^\prime = \frac{f^2}4 \left|\partial_\mu \Sigma \right|^2  &\supset& |\partial_\mu h^+|^2-\frac1{48f^2} \left| h^-\partial_\mu h^+ -h^+\partial_\mu h^-\right|^2 
+ \frac1{1440f^4}\left| h^-\partial_\mu h^+ -h^+\partial_\mu h^-\right|^2 |h^+|^2 \nonumber \\
&& - \frac{1}{80640f^6}\left| h^-\partial_\mu h^+ -h^+\partial_\mu h^-\right|^2 |h^+|^4  + \cdots \ .
\eea
${\cal L}_2$ and ${\cal L}_2^\prime$ look different, but they are not, as we will see later.

\section{Adler's zero condition and shift symmetry}

A few years prior to CCWZ, Adler derived a condition on pion scattering amplitudes as implied by the partially conserved axial current (PCAC) hypothesis, which states that the amplitude for the emission of a single soft pion vanishes \cite{Adler:1964um}. (See also Ref.~\cite{ArkaniHamed:2008gz} for a recent discussion.) This "Adler's zero" condition forbids a constant term in the scattering amplitudes of NGBs and is often loosely described as the NGB only has derivative interactions.

For simplicity let's consider a single flavor of NGB for now. In the $n$-point scattering amplitude of NGBs, using the convention that all momenta are outgoing, the Adler's zero condition together with the Bose symmetry imply the following form:
\bea
{\cal A}(\pi\pi\cdots\to \pi\pi\cdots) &\propto& (p_1+p_2+\cdots +p_n)^2 + {\cal O}(p^4) \nonumber \\
           &\propto& {\cal O}(p^4) \ .
\eea
After imposing the conservation of momentum, $\sum_i p_i = 0$, self-interactions of a single flavor NGB must contain at least four derivatives. The effective lagrangian satisfying the above property is simple and well-known,
\be
{\cal L} = \frac12 \partial_\mu \pi \partial^\mu \pi + {\cal O}(\partial^4)\ .
\ee
While the above effective lagrangian can be derived trivially using the apparatus of CCWZ, a slightly more direct way is to impose a constant shift symmetry on the NGB,
\be
\pi \to \pi +\epsilon \ ,
\ee
where $\epsilon$ is an arbitrary constant. From the CCWZ perspective, the shift symmetry results from the action of the NGB under an element $g=e^{i\epsilon}$ of the  broken $U(1)$ group,
\be
g \ e^{i\pi} = e^{i(\pi+\epsilon)} \ , \qquad g\in U(1)\ .
\ee
The above equation is simply Eq.~(\ref{eq:gxixiu}) when $G=U(1)$. However, it is clear  Eq.~(\ref{eq:gxixiu})  will become quite complicated for a less trivial symmetry breaking pattern.\footnote{In fact, CCWZ avoids computing the transformation property of the NGB completely by dealing with the Cartan-Maurer one form in Eq.~(\ref{eq:cartan}). }  In particular, the CCWZ perspective requires specifying the broken group $G$ in the UV.

\section{Shift Symmetry for two NGBs}
\label{sect:u1}

The shift symmetry in a single NGB arising from $G=U(1)$ is the simplest possibility of nonlinearly realized symmetry. In this section we extend the shift symmetry approach to the next-to-simplest case of two NGBs forming a complex scalar, $\phi=(\pi^1+i\pi^2)/\sqrt{2}$, charged under a $U(1)$ subgroup of the unbroken group: $H \supset U(1)$. However, instead of going  the route of CCWZ by specifying the symmetry breaking pattern in the UV, we will start from the IR by imposing the Adler's zero condition for scattering amplitudes of NGBs of the same flavor:\footnote{The Adler's zero condition is an IR statement because it is a statement on soft particles.}
\bea
&&{\cal A}(\pi^1\pi^1\cdots\to \pi^1\pi^1\cdots) = {\cal O}(p^4) \ ,\\ 
&&{\cal A}(\pi^2\pi^2\cdots\to \pi^2\pi^2\cdots) = {\cal O}(p^4) \ , 
\eea
These two conditions simply state that, when turning off one of the NGBs, the general shift symmetry must reduce to $\pi^i \to \pi^i + \epsilon^i$, $i=1,2$, where $\epsilon^i$ are two arbitrary constants.

The only other condition we will impose, in addition to the Adler's zero condition, is that $U(1)$ symmetry be realized linearly, since it is part of the unbroken group $H$, which suggests the two arbitrary constants are combined into $\epsilon=(\epsilon^1+i\epsilon^2)/\sqrt{2}$. Then, working to linear order in $\epsilon$, the form of the general shift symmetry is completely fixed,
\be
\label{eq:phit}
\phi \mapsto \phi'=\phi + \epsilon -\frac{1}{f^2} (\phi^* \epsilon-\epsilon^* \phi) \phi \ \left[\ \sum_{n=0}^\infty \frac{a_n}{f^{2n}}(\phi^*\phi)^n\right]  \ ,
\ee
where $a_n$ are numerical constants to be determined later and $f$ is the NGB decay constant. When turning off one of the NGBs, say $\pi^1=0$, the general shift symmetry reduces to 
\be
\label{eq:u1vanish}
\left.\phi\right|_{\pi^1=0} \to \left.\phi\right|_{\pi^1=0} + \left.\epsilon \right|_{\epsilon^1=0}\ ,
\ee
which guarantees that the Adler's zero condition is fulfilled for $\pi^2$, and vice versa.

To construct the effective lagrangian for the NGBs,   we look for the NGB covariant derivative ${\cal D}_\mu\phi$ which transforms by a $\phi$- and $\epsilon$-dependent $U(1)$ phase under the action of the general shift symmetry,
\be
\label{eq:dphiuu}
{\cal D}_\mu \phi  \mapsto {\cal D}_\mu \phi'=e^{i  \,\alpha u(\phi, \epsilon)/f} \ {\cal D}_\mu \phi \ ,
\ee
where $\alpha$ is the $U(1)$ charge carried by $\phi$. The forms of ${\cal D}_\mu \phi$ and $u(\phi,\epsilon)$ are also fixed by the Adler's zero condition and the linearly realized $U(1)$ symmetry,
\bea
\label{eq:dphit}
{\cal D}_\mu \phi &=& \partial_\mu \phi - \frac{1}{f^2} (\phi\, \partial_\mu \phi^* -  \partial_\mu \, \phi \phi^*)\phi \left[\ \sum_{n=0}^\infty \frac{b_n}{f^{2n}}(\phi^*\phi)^n \right]  \ , \\
\label{eq:uform}
u(\phi, \epsilon) &=& \frac{1}{f} (\epsilon^* \phi- \phi^* \epsilon)  \left[\sum_{n=0}^\infty \frac{c_n}{f^{2n}}(\phi^*\phi)^n \right]  \ ,
\eea
where $b_n$ and $c_n$ are  numerical constants. Again when setting $\pi^1=0$, ${\cal D}_\mu \to \partial_\mu \pi^2$ and $u\to 0$, which ensure the Adler's zero condition for $\pi^2$ is satisfied.

The numerical constants $a_n, b_n,$ and $c_n$ can be solved order by order in $1/f$ by plugging  Eqs.~(\ref{eq:phit}), (\ref{eq:dphit}) and (\ref{eq:uform}) into the transformation law for ${\cal D}_\mu$ in Eq.~(\ref{eq:dphiuu}). It turns out doing so allows one to determine all coefficients in terms of one single parameter: $a_0$. For example, the first five coefficients are
\bea
(a_0, a_1, a_2, a_3, a_4)&=& (a_0,\ \frac25 a_0^2 ,\ \frac8{35} a_0^3,\  \frac{24}{175} a_0^4,\  \frac{32}{385} a_0^5 ) \nonumber \\
(b_0, b_1, b_2, b_3, b_4)&=& (-\frac12 a_0,\ \frac3{20} a_0^2 ,\ -\frac3{140} a_0^3,\  \frac{1}{560} a_0^4,\  -\frac{3}{30800} a_0^5 ) \nonumber \\
(c_0, c_1, c_2, c_3, c_4)&=& \frac1{\alpha}(-\frac{3i}2 a_0,\ -\frac{3i}4 a_0^2 ,\ -\frac{9i}{20} a_0^3,\  -\frac{153i}{560} a_0^4,\  -\frac{93i}{560} a_0^5 ) \nonumber \ .
\eea
In the end it is possible to obtain a compact expression for the NGB covariant derivative, to all orders in $1/f$,
\be
\label{eq:dphitfix}
{\cal D}_\mu \phi  = \partial_\mu \phi + \phi \frac{\partial_\mu \phi^*\, \phi-\partial_\mu\phi\, \phi^*}{2|\phi|^2}\left(1-\frac{{f}}{|\phi|}\sin\frac{|\phi|}{{f}} \right) \ ,
\ee
where $|\phi|=\sqrt{(\pi^1)^2+(\pi^2)^2}$. In the above the only unknown coefficient $a_0$ has been absorbed into the definition of  the decay constant, ${f}\to  f/\sqrt{6a_0}$. In other words, $a_0$ reflects the arbitrariness in the normalization of the decay constant $f$.

The two-derivative effective lagrangian from Eq.~(\ref{eq:dphitfix}) is
\bea
\label{eq:u1Lag}
{\cal L}^{(2)}& =& {\cal D}_\mu \phi {\cal D}^\mu \phi = \partial_\mu\phi^*\partial^\mu\phi -\frac{|\partial_\mu \phi^*\, \phi-\partial_\mu\phi\, \phi^*|^2}{4|\phi|^2}
\left(1-\frac{{f}^2}{|\phi|^2} \sin^2\frac{|\phi|}{{f}} \right) 
\eea
So we have managed to derive the effective lagrangian for a complex NGB charged under an unbroken $U(1)$ subgroup, using only the IR properties: 1) invariance under $H$ and 2) the Adler's zero condition. No symmetry breaking pattern is specified and, as such, Eq.~(\ref{eq:u1Lag}) is universal among all possible cosets $G/H$. Moreover, working to the first order in $\epsilon$ is sufficient to derive the effective lagrangian.

Going back to the two seemingly different  effective lagrangians for a complex NGB in the Section \ref{sec:ccwz}, it is now easy to check that the only difference between the effective lagrangians from $SU(2)/U(1)$ and $SU(5)/SO(5)$ for the complex NGB is the normalization of the decay constant $f$. Indeed, after a rescaling of $f\to 4f$, ${\cal L}_2$ in Eq.~(\ref{eq:su2u1}) gives precisely  ${\cal L}_2^\prime$ in Eq.~(\ref{eq:su5so5}).

In addition to the NGB covariant derivative ${\cal D}_\mu \phi$, one could work out the associated gauge field ${\cal E}_\mu(\phi)$ in a similar fashion. Specifically, ${\cal E}_\mu$ transforms non-homogeneously and like a gauge field under $U(1)$:
\be
\label{eq:Efield}
{\cal E}_\mu \mapsto  e^{-i u}{\cal E}_\mu e^{i u} - i e^{-i u}\partial_\mu e^{i u} = {\cal E}_\mu+ \partial_\mu u(\phi,\epsilon) \ .
\ee
Again postulating ${\cal E}$ to be of the following form:
\be
{\cal E}_\mu =\frac1{f^2}(\phi^*\partial_\mu \phi - \partial_\mu\phi^* \, \phi) \left[\ \sum_{n=0}^\infty \frac{d_n}{f^{2n}}(\phi^*\phi)^n \right] \ .
\ee
Then solving for Eq.~(\ref{eq:Efield}) order by order in $1/f$ we obtain the following compact expression:
\be
\label{eq:u1efield}
{\cal E}_\mu =  \frac{i}{\alpha} \frac{\partial_\mu \phi^*\, \phi-\partial_\mu\phi\, \phi^*}{|\phi|^2}\sin^2 \frac{|\phi|}{2\tilde{f}} \ ,
\ee
where, as a reminder, $\alpha$ is the charge of $\phi$ under the unbroken $U(1)$.\footnote{In the case of $SU(2)/U(1)$, using the normalization ${\rm Tr}(X^aX^b)=\delta_{ab}$, it is easy to check that $\alpha=-\sqrt{2}$ gives the correct ${\cal E}_\mu$ arising from the coset $SU(2)/U(1)$, for $\phi$ has the charge $-\sqrt{2}$ under the $U(1)$ represented by $T=\sigma^3/\sqrt{2}$.}

With ${\cal D}_\mu\phi$ and ${\cal E}_\mu$, the full effective lagrangian for the complex NGB can be constructed, without ever specifying  a UV coset $G/H$. Moreover, from the IR viewpoint nothing forces the decay constant ${f}$ to be real. If an imaginary decay constant ${f}\to i {f}$ is chosen, one obtains the effective lagrangian for a non-compact coset.

\section{The general case for a simple group}
\label{sect:general}

In this section we generalize the shift symmetry approach to a general unbroken group $H$, taken to be a simple Lie group, whose group generators are $\{T^i, i=1,2,\cdots \}$. Consider a set of scalars, $\pi^a$, furnishing a linear representation of $H$ such that, under the infinitesimal group transformations,
\be
\label{eq:lieH}
\pi^a(x) \to \pi^a(x) + i \alpha^i (T^i)_{ab} \pi^b(x) + {\cal O}(\alpha^2) \ ,
\ee
where $\alpha^i$ are a set of real parameters and $(T^i)_{ab}$ is the corresponding representation of the group generators. A very basic statement, which nonetheless is worth emphasizing, is the behavior of $\pi^a$ under an infinitesimal $H$-rotation is completely characterized by the action of the group generators $T^i$ on $\pi^a$.

For unitary representations the generators $T^i$s are hermitian, $(T^i)^\dagger = T^i$, which we assume. It will be convenient to adopt a basis for the generators such that all generators are purely imaginary and anti-symmetric, $(T^i)^T=-T^i$ and $(T^i)^{*}=-T^i$. For real representations, $\pi^a$ are real fields and  this requirement follows automatically from unitarity of the representation. For  for a set of complex scalars $\phi^a(x)$ furnishing a complex representation $R$, we can write $\pi^a(x) = ( {\rm Re}\ \phi^a,  {\rm Im}\ \phi^a)$ and the generators are
\be
\label{eq:antiti}
T^i = \left( \begin{array}{cc}
          \phantom{-}  i\ {\rm Im}\ T^i_R \ & i\ {\rm Re}\ T^i_R \\
          -  i\ {\rm Re}\ T^i_R\ &   i\  {\rm Im}\ T^i_R 
            \end{array}
            \right)  \ .
            \ee
Notice that the hermiticity of $T_R$ implies ${\rm Re}\, T_R$ is a symmetric matrix while ${\rm Im}\, T_R$ is an anti-symmetric matrix, from which the anti-symmetricity of $T^i$ follows.

\subsection{The Closure Condition}

We would like to consider a set of shift symmetries acting on the $\pi^a$ such that, at leading order, $\pi^a \to \pi^a + \epsilon^a + {\cal O}\left(1/{f^2}\right)$,
for arbitrary real parameters $\epsilon^a$. Our goal is to derive a {\em Closure Condition} such that the effect of an infinitesimal nonlinear action on $\pi^a$ can be compensated by  a field-dependent $H$-rotation,
\be
\label{eq:rst}
|\pi\rangle \to |\pi^{\prime}\rangle \approx |\pi\rangle + |\epsilon\rangle + i \,\alpha^i(\epsilon, \pi)\, |T^i \pi\rangle \ ,
\ee
where we have used the bra-ket notation and  $|T^i \pi\rangle \equiv T^i|\pi\rangle$. Notice the parameter $\alpha^i$ now is dependent on $\epsilon$ and $\pi^a$. Loosely speaking,  Eq.~(\ref{eq:rst}) states that the action of any nonlinearly realized symmetry cannot take $\pi^a$ outside of the representation of $H$.

We are only interested in an infinitesimal nonlinear transformation, implying we only work to the first order in $\epsilon$.  Thus $\alpha^i$ should be linear in $\epsilon$. Adler's zero condition then requires that $\alpha^i(\epsilon, \pi)$ vanishes when reducing to the case of a single NGB by, for example, setting all but $\pi^1$ and $\epsilon^1$ to zero, much like Eq.~(\ref{eq:u1vanish}) in the example of $H=U(1)$. A simple ansatz that realizes the Adler's zero condition is,
\be
\label{eq:1overf2}
\alpha^i = \frac1{f^2}\ \langle \pi T^i \epsilon \rangle +{\cal O}\left(\frac{1}{f^4}\right)\ .
\ee
The general shift at order $1/f^2$ is then
\be
\label{eq:pishiftf4}
|\pi\rangle \to|\pi^\prime\rangle=|\pi\rangle + |\epsilon\rangle + \frac{A_1}{f^2}\ |T^i \pi\rangle  \langle \pi T^i \epsilon \rangle \ ,
\ee
where $A_1$ is a constant to be determined later. Next we construct  NGB covariant derivative ${\cal D}\pi$ which transforms under the shift symmetry in Eq.~(\ref{eq:pishiftf4}) by a pure field-dependent $H$-rotation,
\be
\label{eq:hrotation}
|{\cal D}\pi\rangle \to |{\cal D}\pi^\prime\rangle= e^{i\, u^i(\pi, \epsilon) T^i/f}| {\cal D}\pi\rangle \ ,
\ee
where, at this order in $1/f$,
\bea
|{\cal D}\pi\rangle &=& |\partial \pi\rangle + \frac{B_1}{f^2}|T^i \pi\rangle  \langle\pi T^i \partial\pi \rangle  \ , \\
u^i(\pi, \epsilon) &=&\frac{C_1}{f} \langle\pi T^i \epsilon \rangle    \ .
\eea
Again $B_1$ and $C_1$ are some numerical constants. The above two equations also realize the Adler's zero condition when setting all but one NGB field to zero. The advantage of choosing a real basis in which the generators are anti-symmetric is now evident; otherwise more structures will be present. From Eq.~(\ref{eq:pishiftf4}) we see, under the nonlinear shift symmetry, ${\cal D}\pi$ transforms as
\bea
|{\cal D}\pi^\prime\rangle&=&|{\cal D}\pi\rangle+  \frac{A_1}{f^2}\left( |T^i \partial\pi\rangle  \langle\pi T^i \epsilon \rangle + |T^i \pi\rangle  \langle\partial\pi T^i \epsilon \rangle \right)
        \nonumber \\
        &&  \quad \qquad+ \frac{B_1}{f^2}\left( |T^i \epsilon\rangle  \langle\pi T^i \partial\pi \rangle + |T^i \pi\rangle \langle \epsilon T^i \partial\pi  \rangle \right)\ .
\eea
Expanding Eq.~(\ref{eq:hrotation}) to linear order in $\epsilon$ now gives
\be
(A_1 + i C_1) |T^i \partial\pi\rangle  \langle\pi T^i \epsilon \rangle +(A_1 - B_1) |T^i \pi\rangle  \langle\partial\pi T^i \epsilon \rangle +  B_1 |T^i \epsilon\rangle  \langle\pi T^i \partial\pi \rangle  = 0 \ ,
\ee
which is equivalent to
\be
\label{eq:tabcd}
(T^i)_{ab} (T^i)_{cd} - \frac{A_1 + i C_1}{B_1} (T^i)_{ac} (T^i)_{db} +\frac{A_1-B_1}{B_1} (T^i)_{ad}(T^i)_{bc} = 0 \ .
\ee
A solution to Eq.~(\ref{eq:tabcd}) does not always exist. However, when a solution does exist, we can contract any pair of indices and use the traceless condition of generators to derive
\be
A_1-2B_1 = 0 \ , \qquad A_1+B_1+i C_1 = 0 \ ,
\ee
giving rise to the solution
\be
\label{eq:b0c0}
B_1 = \frac{A_1}2 \ , \qquad C_1 = -\frac32 i A_1 \ .
\ee
In turn, Eq.~(\ref{eq:tabcd}) now becomes
\be
\label{eq:symrep}
(T^i)_{ab} (T^i)_{cd} + (T^i)_{ac} (T^i)_{db} + (T^i)_{ad}(T^i)_{bc} = 0 \ .
\ee
This is the {\em Closure Condition} we set out to look for. When it is satisfied for a given representation of $H$, one can use the general shift symmetry to derive the effective lagrangian, at least to the order of $1/f^2$.  

An important point to stress is that the Closure Condition comes about by requiring the Adler's zero condition and that effects of the shift symmetry can be compensated by a $\pi$-dependent $H$-rotation. The unknown coefficients $B_1$ and $C_1$ are solved in terms of $A_1$ without explicitly specifying the unbroken group $H$, since the solvability only hinges on the Closure Condition. As such, the solution in Eq.~(\ref{eq:b0c0}) is universal and does not depend on the explicit choice of $H$ (other than the requirement that it is a simple group.) This is an important observation, since it implies that, alternatively, we can evaluate the coefficients numerically on an explicit choice of $H$, such as $U(1)\approx SO(2)$. At higher orders in $1/f$, this is our strategy to derive the effective lagrangian.

It turns out the Closure Condition in Eq.~(\ref{eq:symrep}) always holds when the representation under consideration can be embedded in a symmetric coset, and is equivalent to the Jacobi identity for broken generators corresponding to $\pi^a$ in the CCWZ approach. More explicitly, in a coset $G/H$, define $X^a$ to be the broken generators furnishing a given representation $R$ of the unbroken group $H$, whose generators are $T^i$. Both $X^a$ and $T^i$ belong to the adjoint representation of $G$ such that 
\be
[T^i, X^a] = i f^{iab} X^b \ .
\ee
If $R$ can be embedded in a symmetric coset, the commutators of $X^a$ and $X^b$ can be written as
\be
[X^a, X^b]= i f^{abi}\, T^i \ .
\ee
Then the Jacobi identity, $[X^a, [X^b, X^c]]+[X^b, [X^c, X^a]]+[X^c, [X^a, X^b]]$, implies
\be
\label{eq:jacobi}
f^{iab} f^{icd} + f^{ibc}f^{iad} + f^{ica}f^{ibd} = 0 \ .
\ee
On the other hand, recall that  in the adjoint representation there is a state corresponding to each generator $|X^a\rangle$. The action of $T^i$ on the state $|X^a\rangle$ in the adjoint representation is simply given by \cite{Georgi:1982jb}
\be
T^i|X^a\rangle  =  |[T^i, X^a]\rangle = -i f^{iba} |X^b\rangle \ ,
\ee
which implies $-if^{iab}$ is nothing but the matrix entry of $T^i$ in the $R$ representation of $H$, and must coincide with the particular form of $T^i$ in the basis defined in  Eq.~(\ref{eq:antiti}),
\be
\label{eq:tifiab}
(T^i)_{ab} = -i f^{iab} \ .
\ee
In the end, we see the condition in Eq.~(\ref{eq:symrep}) is identical to the Jacobi identity in Eq.~(\ref{eq:jacobi}), when the representation $R$ can be embedded in a symmetric coset. Notice that this is a weaker condition than requiring that $R$ only resides in a symmetric coset. There could be cosets containing $R$ that are non-symmetric, and in these cases the "non-symmetricity" of the coset arises from other sectors not containing $R$.

\subsection{Bootstrapping}
\label{sect:boot}

Going beyond $1/f^2$, we need to consider corrections that are more and more important as the fluctuations in the NGB become comparable to the mass scale $f$. In other words, starting from the neighborhood where $\delta \pi^a \ll f$, we would like to reach the configuration where $\delta \pi^a \sim f$.  Assuming the nonlinearly realized symmetry is smooth and continuous, there are an infinite number of ways to achieve this goal. One possibility is to go in a fixed direction by repeatedly making infinitesimal fluctuations. This bootstrapping procedure is the idea behind the "exponential parameterization" of Lie group that are commonly employed in physics:
\be
e^{i \alpha^i T^i} = \lim_{n\to \infty} \left(1+ i\, \alpha^i T^i/n\right)^n \ ,
\ee
The continuity and smoothness of the group guarantees that, by applying an infinite number of infinitesimal  fluctuations, we can reach the finite  configuration where $\delta \pi^a \sim f$. 

In the particular context here, we are interested in reaching the finite nonlinear transformations by successively applying the infinitesimal nonlinear transformation. It helps to write the infinitesimal  result in Eq.~(\ref{eq:pishiftf4}) as
\be
\label{eq:pinon}
|\pi\rangle \to|\pi^\prime\rangle=|\pi\rangle + \left(1+\frac{A_1}{f^2}|T^i\pi\rangle \langle \pi T^i|\right) |\epsilon\rangle \ ,
\ee
which makes it clear that $1+|T^i\pi\rangle \langle \pi T^i|/f^2$ characterize the infinitesimal  fluctuations in the direction of the nonlinear symmetry. By bootstrapping, the higher order corrections in $1/f$ are parameterized by, schematically,
\be
\lim_{n\to \infty} \left(1+\frac{1}{f^2}|T^i\pi\rangle \langle \pi T^i|\right)^n |\epsilon\rangle  \sim |\epsilon\rangle + \frac1{f^2} |T^i\pi\rangle \langle \pi T^i \epsilon\rangle + \frac1{f^4} |T^i\pi\rangle \langle \pi T^i T^j\pi\rangle \langle \pi T^j \epsilon\rangle +\cdots \ .
\ee
By comparing with Eq.~(\ref{eq:rst}), this parameterization also makes it clear that all higher order nonlinear transformation amounts to compensating the shift in $\epsilon$ by a $\pi$-dependent $H$ rotation.

Given the above considerations, we  parameterize the higher order $1/f$ corrections by
\bea
|\pi^\prime\rangle &=& |\pi\rangle +\sum_{n=0}^{\infty}\frac{A_n}{f^{2n}}\left(\,|T^i \pi\rangle\langle\pi T^i|\,\right)^n |\epsilon\rangle \ , \\
|{\cal D}\pi^\prime\rangle &=&\sum_{n=0}^{\infty}\frac{B_n}{f^{2n}}\left(\,|T^i \pi\rangle\langle\pi T^i|\,\right)^n |\partial \pi\rangle \ , \\
u^i(\pi, \epsilon) &=&\langle\pi T^i |\, \sum_{n=1}^{\infty}\frac{C_n}{f^{2n-1}} \left(\,|T^i \pi\rangle\langle\pi T^i|\,\right)^n\, | \epsilon \rangle \ .
\eea
We have shown that the Closure Condition allows one to solve for $B_1$ and $C_1$ in terms of $A_1$, without  having to specify an unbroken group $H$. Based on the continuity and smoothness assumptions of the nonlinear symmetry, we expect that the Closure Condition is sufficient to ensure a unique solution exists for all the numerical constants $A_n$, $B_n$ and $C_n$.  Indeed, at order $1/f^4$, by multiplying the Closure Condition in Eq.~(\ref{eq:symrep}) by $(T^j)_{de}$ and summing over the index $d$, one arrives at
\be
(T^i)_{ab}(T^iT^j)_{ce} + (T^i)_{bc}(T^iT^j)_{ae} + (T^i)_{ca}(T^iT^j)_{be} =0 \ ,
\ee
which allows one to derive the following linear equations for the coefficients $A_2$, $B_2$ and $C_2$,\footnote{The derivation is somewhat tedious but straightforward. In particular, it is worth keeping in mind that $\langle \pi T^iT^j\pi\rangle = \langle \pi T^{\{i}T^{j\}}\pi\rangle$.}
\bea
&&A_2+B_1A_1-4B_2=0 \ ,\\
 && 2A_2-B_1A_1+2B_2+iC_1B_1=0 \ , \\
 && A_2+B_1A_1+B_2-i(C_2+C_1B_1) = 0 \ . 
\eea
Together with the solution at order $1/f^2$ from Eq.~(\ref{eq:b0c0}), we can solve for all three coefficients in terms of $A_1$,
\be
A_2 = -\frac15 A_1^2 \ , \qquad B_2 = \frac3{40}A_1^2 \ , \qquad C_2 = \frac{3i}{8} A_1^2 \ ,
\ee
again without specifying either the broken group $G$ in the UV or the unbroken group $H$ in the IR. Therefore it is clear now that these coefficients are only there to ensure the closure of nonlinear symmetry with the unbroken $H$ symmetry. As such, they are universal and completely independent of the details of the underlying symmetry groups, both the broken and the unbroken ones!

\subsection{Universal Formulas for CCWZ lagrangian}

As we go to higher and higher orders in $1/f$, the above procedure gets quite cumbersome. However, the universality of the NGB interactions allows us to evaluate the higher order terms explicitly using the simplest possible unbroken group: the $U(1)\approx SO(2)$. Following the procedure in Section \ref{sect:boot}, we derive the following simple result for the NGB covariant derivative 
\be
|{\cal D}\pi\rangle = \left( \frac{\sin \sqrt{{\cal T}}}{\sqrt{\cal T}}\right) |\partial \pi\rangle \ , \qquad
{\cal T}= -3 \frac{A_1}{f^2} |T^i\pi\rangle \langle \pi T^i|  \ .
\ee
Expanding in power series in ${\cal T}$,
 \be
|{\cal D}\pi\rangle =  |\partial \pi\rangle +   \frac{A_1}{2f^2}  |T^i\pi\rangle\langle \pi T^i \partial \pi\rangle + \frac{3A_1^2}{40f^4} |T^i\pi\rangle\langle \pi T^i T^j \pi \rangle \langle \pi  T^j \partial \pi\rangle +\cdots  \ ,
\ee  
which agree with the previous results we obtained. It should be clear by now that the only unknown coefficient $A_1$ is reflecting the arbitrariness in the normalization of the scale $f$, which is not determined from the IR. This is the only information sensitive to the particular symmetry breaking pattern $G/H$, as was seen from the examples of $U(1)$ NGB in Section \ref{sect:u1}. So we may as well re-scale by $f \to \sqrt{-3A_1} f$  so that the NGB covariant derivative is
\be
\label{eq:dpigen}
|{\cal D}\pi\rangle = \left( \frac{\sin \sqrt{{\cal T}}}{\sqrt{\cal T}}\right) |\partial \pi\rangle \ , \qquad
{\cal T}=  \frac{1}{f^2} |T^i\pi\rangle \langle \pi T^i|  \ .
\ee
The associated gauge field ${\cal E}^i$ can be worked out similarly,
\be
{\cal E}^i = \frac{\tilde{D}}{f^2} \langle \partial\pi | \sum_{n=0}^{\infty}\frac{D_n}{f^{2n}} \left( \, |T^i\pi\rangle \langle \pi T^i | \,\right)^n |T^j\pi\rangle \ ,
\ee
where $\tilde{D}$ and $D_n$ are numerical constants. ${\cal E}^i$ transforms like a gauge field under the nonlinear transformation,
\be
{\cal E}^i T^i \to U ({\cal E}^i T^i ) U^{-1} - (\partial U) U^{-1} \ , \quad U = e^{i u^i(\epsilon, \pi) T^i/f} \ .
\ee
In the end, after a similar rescaling of $f\to \sqrt{-3A_1} f$, we obtain
\be
\label{eq:efield}
{\cal E}^i  = \frac{2i}{f^2} \langle \partial\pi |\, \frac{1}{\cal T} \sin^2 \frac{\sqrt{\cal T}}{2}\, |T^i \pi\rangle\ , \qquad {\cal T}=  \frac{1}{f^2} |T^i\pi\rangle \langle \pi T^i| \ .
\ee
In component form, the operator ${\cal T}$ has the matrix entries
\be
({\cal T})_{ab} = \frac1{f^2} (T^i)_{ar} (T^i)_{sb} \pi^r \pi^s \ ,
\ee
where $(T^i)_{ab}$ is the matrix representation of the Lie algebra of $H$, written in the basis we have chosen, and is completely determined in the IR once the choice of the representation and $H$ are made. Eqs. (\ref{eq:dpigen}) and (\ref{eq:efield}) allows one to reconstruct the full CCWZ lagrangian for a given linear representation of $H$ that satisfy the Closure Condition, without recourse to the coset construction.

For $H=U(1)$, Eqs.~(\ref{eq:dpigen}) and (\ref{eq:efield}) agree with our earlier results in Eqs.~(\ref{eq:dphitfix}) and (\ref{eq:u1efield}), respectively, up to a rescaling of the decay constant $f$.  It is worth commenting that, as it stands, Eqs.~(\ref{eq:dpigen}) and (\ref{eq:efield})  are valid only in the particular basis we choose for the generators $T^i$. However, the effective lagrangian constructed out of $|{\cal D}\pi\rangle$, such as the leading two-derivative lagrangian,
\be
{\cal L}^{(2)} = \frac12 \langle {\cal D} \pi | {\cal D}\pi\rangle \ ,
\ee
is basis-independent.

In some examples we considered, such as the fundamental representations of $SO(N)$, the general expressions in Eqs.~(\ref{eq:dpigen}) and (\ref{eq:efield}) can be further simplified. In these cases, we find ${\cal T}^n \sim \langle \pi|\pi\rangle^{n-1} {\cal T}$ and, consequently,
\bea
\label{eq:soND}
|{\cal D}\pi\rangle &=& |\partial \pi\rangle + \frac{2}{\langle\pi|\pi\rangle} \left(1-\frac{f}{\sqrt{2\langle\pi|\pi\rangle}} \sin \frac{\sqrt{2 \langle\pi|\pi\rangle}}{f} \right) \langle \partial  \pi T^i \pi\rangle |T^i \pi\rangle \ , \\
{\cal E}^i &=& \frac{4i}{\langle\pi|\pi\rangle} \sin^2\left(\frac{\sqrt{\langle\pi|\pi\rangle}}{2\sqrt{2}f}\right) \langle \partial \pi T^i \pi\rangle \ .
\eea
Notice that the $SO(2)$ case is the same in the $U(1)$, while the $SO(3)$ fundamental is the same as the adjoint representation of $SU(2)$. The above simplifications, however, do not occur for adjoint representations of either $SO(N)$ or $SU(N)$.

\section{Conclusion and Discussions}
\label{sect:conclusion}

In this work we generalized the shift symmetry to the case of multiple NGBs furnishing a linear representation of a simple Lie group $H$. By requiring the Adler's zero condition and that the nonlinear shift symmetry does not take the NGBs outside of the representation of $H$, we derive at the leading order in $1/f$ a Closure Condition which is sufficient to allow us to reconstruct the full CCWZ lagrangian using only generators of $H$. The knowledge of the broken group $G$ is not necessary. The Closure Condition turned out be equivalent to the requirement that the linear representation under consideration can be embedded in a symmetric coset.

CCWZ is  a top-down approach, by dictating  the broken group $G$ in the UV from the very beginning, while the approach of nonlinear shift symmetry  is entirely bottom-up, working only with the unbroken group $H$ in the IR. The equivalence of the two perspectives indicates that interactions of NGBs, for a given representation of a particular $H$, are universal in the IR and  not sensitive to details of symmetry breaking in the UV. In fact, it seems that the nonlinear NGB interactions  only serve to enforce the Adler's zero condition under the constraint of linearly realized symmetry governed by $H$, at least for those representations that can be embedded in a symmetric coset. The only free parameter in the NGB covariant derivatives and the associated gauge field is the normalization of the decay constant $f$.

While we stressed the universal features of the self-interactions of NGBs from the IR perspective, it is worth noting that one feature of the low-energy effective lagrangian does depend on the broken group $G$ in the UV, that is the number of NGBs. The universality of NGB interactions applies to those furnishing a linear representation of $H$.  This is evident in the specific examples discussed in Sect.~\ref{sec:ccwz}, where one sees self-interactions of a complex NGB charged under an unbroken $U(1)$ in both $SU(2)/U(1)$ and $SU(5)/SO(5)$ differ only by a re-scaling of the decay constant $f$. The number of NGBs in the two cosets are obviously not the same, but the property of the NGB self-interactions within a linear representation of the unbroken group $H$ does not depend on the number of NGBs, which may be viewed as a consequence of the universality.

When the decay constant $f$ is a real number, our general expressions give alternating signs for the higher order corrections in $1/f$, which agree with the finding in Ref.~\cite{Low:2009di} using prime principles such as the unitarity of scattering amplitudes. However,  from the shift symmetry viewpoint, nothing constrains the decay constant to be a real number. When it is an imaginary number, the resulting expressions correspond to a non-compact coset.

From the CCWZ perspective, our results are rather surprising. But the equivalence of the UV and the IR viewpoints implies there must be a way to see that the NGB interactions are independent of $G$ from the top down. The resolution lies in Eq.~(\ref{eq:tifiab}), which identifies $(T^i)_{ab}$, the matrix entry of group generators of $H$, with $-if^{iab}$, the structure constants of the broken group $G$. Indeed, it is not difficult to see that, in CCWZ, the NGB covariant derivatives and the associated gauge fields depend only on $f^{jab}$, when the representation can be embedded in a symmetric coset such that $[X^a, X^b]\sim T^i$. 

There are many representations that can be embedded in a symmetric coset, including the fundamental representations of $SO(N)$ as well as the adjoint representations of any Lie group $G$. The former can be embedded in $SO(N+1)/SO(N)$ while the later in $G\times G/G$. For $SU(N)$ group, however, its fundamental representations cannot be embedded in a symmetric coset. In other words, $SU(N+1)/SU(N)$ is not a symmetric coset. In these cases, one can enlarge the unbroken group from $SU(N)$ to $SU(N)\times U(1)$, which is semi-simple, and the resulting coset $SU(N+1)/SU(N)\times U(1)$ is then symmetric.

From the perspective of the shift symmetry, it is simple to check that generators in the fundamental representations of $SO(N)$  satisfy the Closure Condition in Eq.~(\ref{eq:symrep}), as do generators from any adjoint representation. On the other hand, generators in the $SU(N)$ fundamental do not satisfy the Closure Condition, when written in the real basis chosen in Eq.~(\ref{eq:antiti}). However, if one allows for an extra $U(1)$ generator, again written in the basis of Eq.~(\ref{eq:antiti}), and  let the overall normalization of the $U(1)$ generator  float, the Closure Condition can be satisfied. The particular case of the fundamental representation of $SU(2)\times U(1)$ is of phenomenological importance, because electroweak symmetry is based exactly on $SU(2)_L\times U(1)_Y$ and the Higgs transforms as a fundamental representation under it. The derivation of the effective lagrangian using shift symmetry in this case will be universal among all models where the Higgs arises as a pseudo-NGB \cite{minimalhiggs}.

For future directions, we believe it will be interesting to use the shift symmetry approach to derive the topological interactions, terms in the lagrangian that transform under the shift by a total derivative, instead of being invariant. In addition, it appears that the shift symmetry can accommodate a non-compact coset, if we allow the NGB decay constant to be an imaginary number. From this perspective, it will be interesting to generalize the shift symmetry to include spontaneous breaking of spacetime symmetries, where the counting of NGBs is different from the breaking of internal symmetries \cite{Low:2001bw}.

\begin{acknowledgments}

The author is grateful for  insightful conversations with Nima Arkani-Hamed.  This work was supported in part by the U.S. Department of Energy under Contracts No. DE-AC02-06CH11357 and No. DE-SC0010143, and was initiated at KITP in Santa Barbara, which is supported by the U.S. National Science Foundation under Grant No. NSF PHY11-25915. Hospitality at the Center for Future High Energy Physics at IHEP in Beijing is acknowledged, where part of this work was completed.

\end{acknowledgments}


\end{document}